\documentclass[english,reprint,tightenlines,eqsecnum,aps,prd,nofootinbib]{revtex4}
\usepackage[latin9]{inputenc}
\usepackage{color}
\usepackage{amsmath}
\usepackage{stmaryrd}
\usepackage{graphicx}
\usepackage{esint}

\makeatletter
\@ifundefined{textcolor}{}
{%
 \definecolor{BLACK}{gray}{0}
 \definecolor{WHITE}{gray}{1}
 \definecolor{RED}{rgb}{1,0,0}
 \definecolor{GREEN}{rgb}{0,1,0}
 \definecolor{BLUE}{rgb}{0,0,1}
 \definecolor{CYAN}{cmyk}{1,0,0,0}
 \definecolor{MAGENTA}{cmyk}{0,1,0,0}
 \definecolor{YELLOW}{cmyk}{0,0,1,0}
}

\makeatother

\usepackage{babel}
\begin{document}

\title{Two-point function of a quantum scalar field in the interior region
of a Reissner-Nordstrom black hole}

\author{Assaf Lanir, Adam Levi, Amos Ori and Orr Sela}

\address{Department of physics, Technion-Israel Institute of Technology, Haifa
32000, Israel}
\begin{abstract}
We derive explicit expressions for the two-point function of a massless
scalar field in the interior region of a Reissner-Nordstrom black
hole, in both the Unruh and Hartle-Hawking quantum states. The two-point
function is expressed in terms of the standard $lm\omega$ modes of
the scalar field (those associated with a spherical harmonic $Y_{lm}$
and a temporal mode $e^{-i\omega t}$), which can be conveniently
obtained by solving an ordinary differential equation, the radial
equation. These explicit expressions are the internal analogs of the
well known results in the external region (originally derived by Christensen
and Fulling), in which the two-point function outside the black hole
is written in terms of the external $lm\omega$ modes of the field.
They allow the computation of $<\Phi^{2}>_{ren}$ and the renormalized
stress-energy tensor inside the black hole, after the radial equation
has been solved (usually numerically). In the second part of the paper,
we provide an explicit expression for the trace of the renormalized
stress-energy tensor of a \emph{minimally-coupled} massless scalar
field (which is non-conformal), relating it to the d'Alembertian of
$<\Phi^{2}>_{ren}$. This expression proves itself useful in various
calculations of the renormalized stress-energy tensor. 
\end{abstract}
\maketitle

\section{introduction}

In the framework of semiclassical general relativity, the gravitational
field is treated classically as a curved spacetime while other fields
are treated as quantum fields residing in this background spacetime.
The relation between the spacetime geometry and the stress-energy
of the quantum fields is described by the semiclassical Einstein equation
\begin{equation}
G_{\mu\nu}=8\pi\left\langle \hat{T}_{\mu\nu}\right\rangle _{ren},\label{semiclassical Einstein}
\end{equation}
where $G_{\mu\nu}$ is the Einstein tensor of spacetime, and $<\hat{T}_{\mu\nu}>_{ren}$
is the renormalized stress-energy tensor (RSET), which is the renormalized
expectation value of the stress-energy tensor operator $\hat{T}$,
associated with the quantum fields. In Eq. \eqref{semiclassical Einstein}
and throughout this paper we adopt standard geometric units $c=G=1$,
and signature $\left(-+++\right)$.

The main challenge in analyzing the semiclassical Einstein equation
is the computation of the RSET. Even when the background geometry
is fixed and the corresponding metric is given, performing a procedure
of renormalization is not an easy task. The task becomes much more
difficult upon trying to solve the full self-consistent problem represented
by Eq. \eqref{semiclassical Einstein}. One reason (besides the obvious
numerical challenge) is that Eq. \eqref{semiclassical Einstein} admits
runaway solutions \cite{Runaway}. 

An example of a quantum field that is often chosen for its simplicity
is that of a scalar field, which satisfies the Klein-Gordon equation
\begin{equation}
\left(\boxempty-m^{2}-\xi R\right)\hat{\Phi}=0,\label{Klein Gordon General}
\end{equation}
where $\hat{\Phi}$ is the scalar field operator, $m$ denotes the
field's mass, and $\xi$ is the coupling constant. As a first stage
towards calculating the RSET, it is customary to begin by calculating
$<\hat{\Phi}^{2}>_{ren}$, as this quantity is endowed with many of
the essential features of the RSET, but is simpler to compute. In
other words, $<\hat{\Phi}^{2}>_{ren}$ serves as a simple toy model
for the RSET. 

The standard method to calculate quantities which are quadratic in
the field and its derivatives is \emph{point-splitting} (or \emph{covariant
point separation}), developed by DeWitt for $<\hat{\Phi}^{2}>_{ren}$
and by Christensen for the RSET \cite{Dewitt - Dynamical theory of groups and fields,Christiansen}.
In this method, one splits the point $x$, in which $<\hat{\Phi}^{2}>_{ren}$
is evaluated, and writes it as a product of the field operators at
two different points, namely the two-point function (TPF) $<\hat{\Phi}\left(x\right)\hat{\Phi}\left(x'\right)>$.
One then subtracts from the TPF a known counter-term and takes the
limit $x'\rightarrow x$, thereby obtaining $<\hat{\Phi}^{2}>_{ren}$. 

The numerical implementation of the limit described above turns out
to be very difficult. To overcome this difficulty, practical methods
to implement the point-splitting scheme were developed by Candelas,
Howard, Anderson and others \cite{WKB outside 1,WKB outside 2,WKB outside 3,WKB outside 4}.
These techniques however all relied on Wick rotation, namely, they
required the background to admit a euclidean sector (usually employing
a high-order WKB approximation for the field modes on this sector).

Recently, a more versatile method to implement the point-splitting
scheme was developed, the \textit{pragmatic mode-sum regularization}
(PMR) scheme. In this method, the background does not need to admit
a euclidean sector, and the WKB approximation is not used. Instead,
the background must only admit a single Killing field. The PMR method
was tested and used to compute $<\hat{\Phi}^{2}>_{ren}$ and the RSET
in the exterior part of a Schwarzschild black hole (BH) \cite{Adam1,Adam2,Adam3,Adam5},
a Reissner-Nordstrom (RN) BH \cite{Adam6}, and a Kerr BH \cite{Adam4}. 

So far most of the calculations of $<\hat{\Phi}^{2}>_{ren}$ and the
RSET were performed on the exterior part of BHs. The only exception
we know of is Ref. \cite{Candelas_and_Jensen}, which calculated $<\hat{\Phi}^{2}>_{ren}$
for the interior part of Schwarzschild, in the Hartle-Hawking state.
The main reason is probably that the internal calculation requires
a longer analytical derivation for expressing the TPF in terms of
the standard Eddington-Finkelstein modes. In addition, the internal
calculation requires numerical computation of modes both inside and
outside the BH. 

Although it is not an easy task, there is a great interest in studying
the semiclassical quantum effects in the interior of BHs. One obvious
reason is the question whether semiclassical effects could resolve
the spacetime singularity inside e.g. a Schwarzschild BH \cite{ref1,ref2}.\textcolor{red}{{}
}Yet another motivation is the quest to understand the fate of the
inner horizon inside spinning or charged BHs, when realistic perturbations
are taken into account. For both spinning and charged BHs, in the
corresponding unperturbed classical solution (the Kerr or RN solution
respectively), the inner horizon is a perfectly smooth null hypersurface.
Classical perturbations typically convert the smooth inner horizon
into a curvature singularity, which is nevertheless null and weak
(i.e. tidally non-destructive). One may expect, however, that semiclassical
energy-momentum fluxes could have a stronger potential effect on the
inner horizon (perhaps converting the latter into a strong, i.e. tidally
destructive, spacelike singularity). A first and important step towards
clarifying this issue is the RSET computation on the background of
the classical RN or Kerr geometry \textemdash{} inside the BH (and
particularly near the inner horizon). 

Several works have been previously made in the attempt to address
this issue analytically, for RN \cite{Inner Horizon 1}, Kerr \cite{Inner Horizon 2},
and even Kerr-Newmann \cite{Inner Horizon 3} background spacetimes.
Generally speaking, these works suggested that indeed the RSET is
likely to diverge at the inner horizon. However, the results obtained
so far (at least in 4D) are still not entirely conclusive. Thus, Ref.
\cite{Inner Horizon 1} analyzed a 2D RN model and found RSET divergence
at the Cauchy horizon (CH). In addition, Refs. \cite{Inner Horizon 1,Inner Horizon 2,Inner Horizon 3}
obtained several relations between various RSET components at the
CH (at leading order) in 4D. They showed that certain nontrivial quantities
have to vanish in order for the RSET to be regular there, which may
suggest (but does not prove) divergence. The strongest CH result was
derived by Hiscock in Ref. \cite{Inner Horizon 3}: He explicitly
showed that in Unruh state in RN, the semiclassical fluxes must diverge
at either the ingoing section (i.e. the CH) or the outgoing section
of the inner horizon \textemdash{} or possibly at both. Still, this
result leaves open the possibility of a perfectly regular RSET at
the CH. A possible way to conclusively address this issue is to explicitly
compute the RSET inside a Kerr and/or RN black hole, and to obtain
its asymptotic behavior on approaching the CH. This, however, requires
the extension of the RSET computation infrastructure to the internal
part of the BH. Here we undertake this goal in the case of RN background
(deferring the more complicated Kerr case to future works). 

More specifically, in this paper we address two different (and perhaps
loosely related) issues. The first one is directly related to the
main objective described above: We consider a massless quantum scalar
field and construct explicit expressions for its TPF in the interior
of a RN black hole, expressing it as a sum of standard Eddington-Finkelstein
modes that are naturally defined in the interior region. Specifically,
we focus on the symmetrized form of the TPF, which is also known as
the Hadamard elementary function
\begin{equation}
G^{\left(1\right)}\left(x,x'\right)=\left\langle \left\{ \hat{\Phi}\left(x\right),\hat{\Phi}\left(x'\right)\right\} \right\rangle ,\label{Hadamard}
\end{equation}
for both the Unruh and the Hartle-Hawking quantum states \footnote{In the point-splitting procedure, it is common to use the Hadamard
function $G^{\left(1\right)}\left(x,x'\right)$ instead of $\left\langle \hat{\Phi}\left(x\right)\hat{\Phi}\left(x'\right)\right\rangle $.}, where $\left\{ ,\right\} $ denotes anti-commutation. The final
expressions can be found in Eqs. \eqref{Hadamard Unruh} and \eqref{Hadamard H-H}.
Using these expressions one can calculate $<\hat{\Phi}^{2}>_{ren}$
(using e.g. the PMR method) after numerically solving the radial equation
(an ordinary differential equation) needed for constructing the standard
Eddington-Finkelstein modes. 

The second issue is pointing out an identity for the RSET of a minimally-coupled
massless scalar field, which is analogous to the well known trace-anomaly
identity for conformal fields. The identity is easily derived using
known expressions, and we found it very useful. Yet we could not find
any explicit indications for it in the literature. 

The general considerations described above motivate one to study the
internal semiclassical effects primarily in quantum states that may
be considered as ``vacuum'', like Hartle-Hawking and Unruh states
(and especially in the latter state, which characterizes the actual
evaporating BHs). Indeed, the results of sections \ref{sec:The-Unruh-state}
and \ref{sec:The-Hartle-Hawking-state} explicitly refer to the TPF
in the Unruh and Hartle-Hawking states respectively. Note, however,
that the result derived in Sec. \ref{sec:trace-anomaly-for} (the
trace of a minimally-coupled scalar field) applies to \emph{any} quantum
state. 

In forthcoming papers, the results of this paper will be used to compute
$<\hat{\Phi}^{2}>_{ren}$ and the RSET in the interior region of Schwarzschild
and RN spacetimes using two different approaches. One approach \cite{Orr-Amos-Preparation}
uses analytical asymptotic approximations for the Eddington-Finkelstein
modes, in order to analyze the leading-order divergence (if it occurs)
of $<\hat{\Phi}^{2}>_{ren}$ and the RSET upon approaching the inner
horizons. The other approach \cite{Assaf-Adam-Amos-Preparation} uses
a numerical computation of the Eddington-Finkelstein modes in the
interior region of the BH. It then utilizes the aforementioned PMR
method to calculate $<\hat{\Phi}^{2}>_{ren}$ in the interior. The
second approach was already used for the Schwarzschild case and the
results agree fairly well with those presented in \cite{Candelas_and_Jensen}.
The two approaches complement each other and allow for cross-checking
the results.

The organization of this paper is as follows. In section \ref{sec:Preliminaries}
we introduce all the necessary preliminaries needed for our derivation.
Then, in sections \ref{sec:The-Unruh-state} and \ref{sec:The-Hartle-Hawking-state},
we develop expressions for the TPF calculated in Unruh and Hartle-Hawking
states, respectively. Section \ref{sec:trace-anomaly-for} analyzes
the trace of a massless, minimal-coupled, scalar field. In section
\ref{sec:Discussion} we discuss our results and possible extensions. 

\section{Preliminaries\label{sec:Preliminaries}}

Before we begin the computation, let us start by defining the coordinate
systems, sets of modes and quantum states which we use in this paper,
and the form of the Hadamard function outside the BH. 

\subsection{Coordinate systems}

In this paper we consider the RN spacetime, which in the standard
Schwarzschild coordinates has the following metric: 
\[
ds^{2}=-\left(1-\frac{2M}{r}+\frac{Q^{2}}{r^{2}}\right)dt^{2}+\left(1-\frac{2M}{r}+\frac{Q^{2}}{r^{2}}\right)^{-1}dr^{2}+r^{2}\left(d\theta^{2}+\sin^{2}\theta d\varphi^{2}\right).
\]
The event horizon $\left(r=r_{+}\right)$ and the inner horizon $\left(r=r_{-}\right)$
are located at the two roots of $g_{tt}=0$, i.e. 
\[
r_{\pm}=M\pm\sqrt{M^{2}-Q^{2}}.
\]
Throughout this paper, only the region $r\geq r_{-}$ will be concerned.
The surface gravity parameters at the two horizons, $\kappa_{\pm}$
, are given by:
\[
\kappa_{\pm}=\frac{r_{+}-r_{-}}{2r_{\pm}^{2}}.
\]

We define the tortoise coordinate, $r_{*}$, both in the interior
and the exterior regions, using the standard relation
\[
\frac{dr}{dr_{*}}=1-\frac{2M}{r}+\frac{Q^{2}}{r^{2}}.
\]
Specifically, we choose the integration constants in the interior
and the exterior regions such that in both regions 
\[
r_{*}=r+\frac{1}{2\kappa_{+}}\ln\left(\frac{\left|r-r_{+}\right|}{r_{+}-r_{-}}\right)-\frac{1}{2\kappa_{-}}\ln\left(\frac{\left|r-r_{-}\right|}{r_{+}-r_{-}}\right).
\]
Note that $r_{+}$ corresponds to $r_{*}\rightarrow-\infty$ (both
for $r_{*}$ defined in the exterior region and for that defined in
the interior) and $r_{-}$ to $r_{*}\rightarrow+\infty$. 

\begin{figure}
\begin{centering}
\includegraphics[scale=0.7]{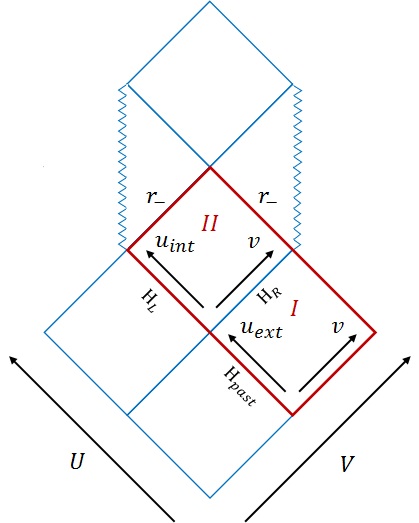}
\par\end{centering}
\caption{Penrose diagram of Reissner-Nordstrom spacetime. In the exterior region
(region $I$), we use the external Eddington-Finkelstein coordinates,
while in the interior (region $II$), we use the internal ones. In
addition, the Kruskal coordinate system is shown and is defined throughout
both regions $I$ and $II$. The red-framed area denotes the region
in the eternal Reissner-Nordstrom spacetime which concerns this paper,
i.e. regions $I$ and $II$. \label{fig:Penrose diag RN}}
\end{figure}

The Eddington-Finkelstein coordinates are defined in the exterior
region by
\[
u_{\mathrm{ext}}=t-r_{*}\quad,\quad v=t+r_{*},\quad\left(\textrm{outside}\right)
\]
while in the interior region they are 
\[
u_{\mathrm{int}}=r_{*}-t\quad,\quad v=r_{*}+t,\quad\left(\textrm{inside}\right).
\]
Note that the $v$ coordinate is continuously defined in both regions
I and II of Fig. \ref{fig:Penrose diag RN}. 

The Kruskal coordinates (corresponding to the event horizon $r_{+}$)
are defined in terms of the exterior and interior Eddington-Finkelstein
coordinates by
\begin{equation}
U\left(u_{\mathrm{ext}}\right)=-\frac{1}{\kappa_{+}}\exp\left(-\kappa_{+}u_{\mathrm{ext}}\right)\quad,\quad V\left(v\right)=\frac{1}{\kappa_{+}}\exp\left(\kappa_{+}v\right),\quad\left(\textrm{outside}\right)\label{Kruskal ext}
\end{equation}
and
\begin{equation}
U\left(u_{\mathrm{int}}\right)=\frac{1}{\kappa_{+}}\exp\left(\kappa_{+}u_{\mathrm{int}}\right)\quad,\quad V\left(v\right)=\frac{1}{\kappa_{+}}\exp\left(\kappa_{+}v\right),\quad\left(\textrm{inside}\right).\label{Kruskal int}
\end{equation}

We make the following notations: $H_{\mathrm{past}}$ denotes the
past horizon {[}i.e. the hypersurface $\left(U<0\,,\,V=0\right)${]},
PNI denotes past null infinity {[}i.e. $\left(U=-\infty\,,\,V>0\right)${]}.
$H_{L}$ is the ``left event horizon'' $\left(U>0\,,\,V=0\right)$,
and $H_{R}$ is the ``right event horizon'' $\left(U=0\,,\,V>0\right)$.
See Fig. \ref{fig:Penrose diag RN}.

\subsection{Modes\label{Modes-and-quantum}}

In this paper we are considering a massless quantum scalar field operator
$\hat{\Phi}$ in RN, satisfying the Klein-Gordon equation
\begin{equation}
\boxempty\hat{\Phi}=0.\label{KG}
\end{equation}
The quantum states considered in this paper are conveniently defined
via a decomposition of this field into sets of modes satisfying Eq.
\eqref{KG} with certain initial conditions. Therefore it is useful
to consider various complete sets of modes in different regions of
spacetime. \footnote{Note that not all of these sets of modes are necessarily related to
the definition of a quantum state. Specifically, the internal ``right''
and ``left'' modes (see below) are introduced merely for mathematical
convenience. } 

First we define the \emph{Unruh modes}, $g_{\omega lm}^{\mathrm{up}}$
and $g_{\omega lm}^{\mathrm{in}}$ (for $\omega>0$). Exploiting the
spherical symmetry, we define these modes by decomposing them in the
following standard way: 
\begin{equation}
g_{\omega lm}^{\Lambda}\left(x\right)=\omega^{-1/2}C_{lm}\left(x\right)\tilde{g}_{\omega l}^{\Lambda}\left(x\right),\label{full Unruh modes}
\end{equation}
where 
\begin{equation}
C_{lm}\left(x\right)=\left(4\pi\right)^{-1/2}\frac{1}{r}Y_{lm}\left(\theta,\varphi\right),\label{C}
\end{equation}
$\varLambda$ denotes ``in'' and ``up'', and $\tilde{g}_{\omega l}^{\Lambda}$
are solutions of the following two-dimensional wave equation, obtained
by substituting Eq. \eqref{full Unruh modes} in Eq. \eqref{KG}:
\begin{equation}
\tilde{g}_{,r_{*}r_{*}}^{\Lambda}-\tilde{g}_{,tt}^{\Lambda}=V_{l}\left(r\right)\tilde{g}^{\Lambda},\label{wave eq}
\end{equation}
where
\begin{equation}
V_{l}\left(r\right)=\left(1-\frac{2M}{r}+\frac{Q^{2}}{r^{2}}\right)\left[\frac{l\left(l+1\right)}{r^{2}}+\frac{2M}{r^{3}}-\frac{2Q^{2}}{r^{4}}\right].\label{Potential}
\end{equation}
The two sets of independent solutions, $\tilde{g}_{\omega l}^{\mathrm{up}}$
and $\tilde{g}_{\omega l}^{\mathrm{in}}$, are defined according to
the following initial conditions: 
\begin{equation}
\tilde{g}_{\omega l}^{\mathrm{up}}=\left\{ \begin{array}{c}
e^{-i\omega U}\quad,\quad H_{\mathrm{past}}\\
e^{-i\omega U}\quad,\quad H_{L}\\
0\quad,\quad\mathrm{\mathrm{PNI}}
\end{array}\right.\qquad,\qquad\tilde{g}_{\omega l}^{\mathrm{in}}=\left\{ \begin{array}{c}
0\quad,\quad H_{\mathrm{past}}\\
0\quad,\quad H_{L}\\
e^{-i\omega v}\quad,\quad\mathrm{\mathrm{PNI}}
\end{array}\right..\label{g-in-up}
\end{equation}
We further introduce for later use the two parts of $\tilde{g}_{\omega l}^{\mathrm{up}}$
which, too, satisfy Eq. \eqref{wave eq} but with initial conditions
\[
\tilde{g}_{\omega l}^{\mathrm{past}}=\left\{ \begin{array}{c}
e^{-i\omega U}\quad,\quad H_{\mathrm{past}}\\
0\quad,\quad H_{L}\\
0\quad,\quad\mathrm{\mathrm{PNI}}
\end{array}\right.\qquad,\qquad\tilde{g}_{\omega l}^{\mathrm{L}}=\left\{ \begin{array}{c}
0\quad,\quad H_{\mathrm{past}}\\
e^{-i\omega U}\quad,\quad H_{L}\\
0\quad,\quad\mathrm{\mathrm{PNI}}
\end{array}\right..
\]
These modes are defined in both regions $I$ and $II$, i.e. throughout
the red-framed area of Fig. \ref{fig:Penrose diag RN}. Note that,
by additivity, $\tilde{g}_{\omega l}^{\mathrm{past}}\left(x\right)+\tilde{g}_{\omega l}^{\mathrm{L}}\left(x\right)=\tilde{g}_{\omega l}^{\mathrm{up}}\left(x\right)$.

Next, we turn to the definition of the \emph{outer Eddington-Finkelstein
modes}, $f_{\omega lm}^{\mathrm{up}}$ and $f_{\omega lm}^{\mathrm{in}}$.
Similar to Eq. \eqref{full Unruh modes} above, we decompose the modes
as:
\begin{equation}
f_{\omega lm}^{\Lambda}\left(x\right)=\left|\omega\right|^{-1/2}C_{lm}\left(x\right)\tilde{f}_{\omega l}^{\Lambda}\left(x\right),\label{full Edding modes}
\end{equation}
and like the functions $\tilde{g}_{\omega l}^{\Lambda}$, the functions
$\tilde{f}_{\omega l}^{\Lambda}$ satisfy Eq. \eqref{wave eq}, but
are defined only in region $I$ (see Fig. \ref{fig:Penrose diag RN}).
The two independent sets of modes that correspond to ``in'' and
``up'' are defined according to the following initial conditions: 

\begin{equation}
\tilde{f}_{\omega l}^{\mathrm{in}}=\left\{ \begin{array}{c}
0\quad,\quad H_{\mathrm{past}}\\
e^{-i\omega v}\quad,\quad\mathrm{PNI}
\end{array}\right.\qquad,\qquad\tilde{f}_{\omega l}^{\mathrm{up}}=\left\{ \begin{array}{c}
e^{-i\omega u_{\mathrm{ext}}}\quad,\quad H_{\mathrm{past}}\\
0\quad,\quad\mathrm{PNI}
\end{array}\right..\label{initial conditions in and up}
\end{equation}
Note that we formally define the modes $f_{\omega lm}^{\Lambda}$
for negative values of $\omega$ as well {[}see Eq. \eqref{full Edding modes}{]},
although our final expressions will contain only modes of positive
values of $\omega$, i.e. positive frequency modes. 

These outer Eddington-Finkelstein\emph{ }modes are especially useful,
as they can be decomposed into radial functions which satisfy an \emph{ordinary}
differential equation and can therefore be easily computed numerically.
The decomposition is as follows:
\begin{equation}
\tilde{f}_{\omega l}^{\mathrm{in}}\left(r,t\right)=e^{-i\omega t}\varPsi_{\omega l}^{\mathrm{in}}\left(r\right)\,\,,\,\,\tilde{f}_{\omega l}^{\mathrm{up}}\left(r,t\right)=e^{-i\omega t}\varPsi_{\omega l}^{\mathrm{up}}\left(r\right).\label{decomp rad ext}
\end{equation}
Substituting these decompositions into Eq. \eqref{wave eq} yields
the following radial equation for $\varPsi_{\omega l}^{\mathrm{\Lambda}}$:
\begin{equation}
\varPsi_{,r_{*}r_{*}}^{\mathrm{\Lambda}}+\left[\omega^{2}-V_{l}\left(r\right)\right]\varPsi^{\mathrm{\Lambda}}=0,\label{radial-exterior}
\end{equation}
where the effective potential $V_{l}$ is given by Eq. \eqref{Potential}.
In terms of the radial functions $\varPsi_{\omega l}^{\mathrm{\Lambda}}$,
the initial conditions given in Eq. \eqref{initial conditions in and up}
are translated to
\begin{equation}
\varPsi_{\omega l}^{\mathrm{in}}\left(r\right)\cong\left\{ \begin{array}{c}
\tau_{\omega l}^{\mathrm{in}}e^{-i\omega r_{*}}\quad,\quad r_{*}\rightarrow-\infty\\
e^{-i\omega r_{*}}+\rho_{\omega l}^{\mathrm{in}}e^{i\omega r_{*}}\quad,\quad r_{*}\rightarrow\infty
\end{array}\right.\label{bc exterior in}
\end{equation}
and 

\begin{equation}
\varPsi_{\omega l}^{\mathrm{up}}\left(r\right)\cong\left\{ \begin{array}{c}
e^{i\omega r_{*}}+\rho_{\omega l}^{\mathrm{up}}e^{-i\omega r_{*}}\quad,\quad r_{*}\rightarrow-\infty\\
\tau_{\omega l}^{\mathrm{up}}e^{i\omega r_{*}}\quad,\quad r_{*}\rightarrow\infty
\end{array}\right.,\label{bc exterior up}
\end{equation}
where $\rho_{\omega l}^{\Lambda}$ and $\tau_{\omega l}^{\varLambda}$
are the reflection and transmission coefficients (corresponding to
the mode $\tilde{f}_{\omega l}^{\Lambda}$), respectively. Solving
numerically Eq. \eqref{radial-exterior} together with the boundary
conditions \eqref{bc exterior in} and \eqref{bc exterior up} yields
$\varPsi_{\omega l}^{\mathrm{\Lambda}}\left(r\right)$, which then
gives the modes $f_{\omega lm}^{\mathrm{in}}$ and $f_{\omega lm}^{\mathrm{\mathrm{\mathrm{up}}}}$
using Eqs. \eqref{decomp rad ext} and \eqref{full Edding modes}. 

In a similar way, we define two sets of \emph{inner Eddington-Finkelstein
modes}, which are similarly decomposed according to Eq. \eqref{full Edding modes},
and the corresponding functions $\tilde{f}_{\omega l}^{\Lambda}$
again satisfy Eq. \eqref{wave eq}. Here, however, $\Lambda$ denotes
``right'' ($R$) and ``left'' ($L$) corresponding to the following
initial conditions on the left and right event horizons: 

\begin{equation}
\tilde{f}_{\omega l}^{\mathrm{L}}=\left\{ \begin{array}{c}
e^{-i\omega u_{\mathrm{int}}}\quad,\quad H_{L}\\
0\quad,\quad H_{R}
\end{array}\right.\qquad,\qquad\tilde{f}_{\omega l}^{\mathrm{R}}=\left\{ \begin{array}{c}
0\quad,\quad H_{L}\\
e^{-i\omega v}\quad,\quad H_{R}
\end{array}\right..\label{initial conditions L and R}
\end{equation}
Note that these modes are defined only in region $II$ (see Fig. \ref{fig:Penrose diag RN}).

As in the external region, the modes $f_{\omega lm}^{\mathrm{L}}$
and $f_{\omega lm}^{\mathrm{R}}$ are useful since they can be decomposed
into a radial function, which can be easily computed numerically.
(Note that in this case we have a single radial function instead of
two.) The decomposition is as follows: 
\begin{equation}
\tilde{f}_{\omega l}^{\mathrm{L}}\left(r,t\right)=\psi_{\omega l}\left(r\right)e^{i\omega t}\,\,,\,\,\tilde{f}_{\omega l}^{\mathrm{R}}\left(r,t\right)=\psi_{\omega l}\left(r\right)e^{-i\omega t}.\label{decomp rad int}
\end{equation}
As before, substituting these decompositions into Eq. \eqref{wave eq}
yields the radial equation \eqref{radial-exterior} for $\psi_{\omega l}$.
In terms of this radial function, the initial conditions given in
Eq. \eqref{initial conditions L and R} reduce to the single condition
\begin{equation}
\psi_{\omega l}\cong e^{-i\omega r_{*}},\,\,\,r_{*}\rightarrow-\infty.\label{bdry cond}
\end{equation}
Then, solving numerically Eq. \eqref{radial-exterior} together with
the initial condition \eqref{bdry cond} yields $\psi_{\omega l}\left(r\right)$,
which in turn gives the modes $f_{\omega lm}^{\mathrm{L}}$ and $f_{\omega lm}^{\mathrm{\mathrm{R}}}$
using Eqs. \eqref{decomp rad int} and \eqref{full Edding modes}. 

\subsection{Quantum states}

One can use the Unruh modes and the outer Eddington-Finkelstein modes
to define the Unruh and Boulware states, respectively. In order to
define the Unruh state, we decompose the scalar field operator in
terms of the Unruh modes $g_{\omega lm}^{\mathrm{up}}$ and $g_{\omega lm}^{\mathrm{in}}$
as follows: 
\begin{equation}
\hat{\Phi}\left(x\right)=\intop_{0}^{\infty}d\omega\sum_{\Lambda,l,m}\left[g_{\omega lm}^{\Lambda}\left(x\right)\hat{a}_{\omega lm}^{\Lambda}+g_{\omega lm}^{\Lambda*}\left(x\right)\hat{a}_{\omega lm}^{\Lambda\dagger}\right].\label{phi decomp Unruh}
\end{equation}
Then the Unruh state $\left|0\right\rangle _{U}$ is defined by \cite{Unruh}
\[
\hat{a}_{\omega lm}^{\Lambda}\left|0\right\rangle _{U}=0,
\]
where $\hat{a}_{\omega lm}^{\Lambda}$ are the annihilation operators
appearing in Eq. \eqref{phi decomp Unruh}. This quantum state describes
an evaporation of a BH, i.e. it involves an outgoing flux of radiation
at infinity (with no incoming waves at PNI). The vacuum expectation
value of the stress-energy tensor in this state is regular at $H_{R}$
(but not at $H_{\mathrm{past}}$) \cite{Candelas}.

Similarly, the Boulware state \cite{Boulware} is defined using the
decomposition 
\begin{equation}
\hat{\Phi}\left(x\right)=\intop_{0}^{\infty}d\omega\sum_{\Lambda,l,m}\left[f_{\omega lm}^{\Lambda}\left(x\right)\hat{b}_{\omega lm}^{\Lambda}+f_{\omega lm}^{\Lambda*}\left(x\right)\hat{b}_{\omega lm}^{\Lambda\dagger}\right]\label{phi decomp Boulware}
\end{equation}
in terms of the outer Eddington-Finkelstein modes, $f_{\omega lm}^{\mathrm{up}}$
and $f_{\omega lm}^{\mathrm{in}}$, and the condition 
\[
\hat{b}_{\omega lm}^{\Lambda}\left|0\right\rangle _{B}=0,
\]
where $\hat{b}_{\omega lm}^{\Lambda}$ are the annihilation operators
appearing in Eq. \eqref{phi decomp Boulware}. The Boulware state
matches the usual notion of a vacuum at infinity, i.e. the vacuum
expectation value of the stress-energy tensor in this state vanishes
at infinity. On the other hand, this vacuum expectation value, evaluated
in a freely falling frame, is singular at the past and future event
horizons \cite{Candelas}. Note that this quantum state is only defined
in region I.

The Hartle-Hawking state \cite{Hartle =000026 Hawking} corresponds
to a thermal bath of radiation at infinity. In this state the Hadamard
function and the RSET are regular on both $H_{\mathrm{past}}$ and
$H_{R}$ \cite{Candelas}. Formally this quantum state may be defined
by an analytic continuation to the Euclidean sector. Here we shall
primarily be interested in the mode structure of the Hadamard function
in this state, given in the next subsection.

\subsection{The Hadamard function outside the black hole\label{Hadamard function outside}}

As shown above in Eq. \eqref{phi decomp Unruh}, the scalar field
operator can be decomposed in terms of the Unruh modes. Substituting
this expression into the definition of the Unruh state Hadamard function,
using Eq. \eqref{Hadamard}, readily yields a mode-sum expression
for this function in terms of the Unruh modes. In the exterior region
of the BH, these Unruh modes can be reexpressed in terms of the outer
Eddington-Finkelstein modes. Then, using this relation, an expression
for the Unruh state Hadamard function in terms of the latter modes
can be obtained. 

This procedure was carried out in \cite{Candelas,Christensen =000026  Fulling},
and we quote here the final result \footnote{In \cite{Candelas,Christensen =000026  Fulling} the results were
given specifically for the Schwarzschild case. Nevertheless, the translation
of these results from Schwarzchild to RN  is straightforward (one
only needs to replace Schwarzschild's surface gravity $\kappa$ by
the corresponding RN parameter $\kappa_{+}$). }:

\begin{equation}
G_{U}^{\left(1\right)}\left(x,x'\right)=\intop_{0}^{\infty}d\omega\sum_{l,m}\left[\coth\left(\frac{\pi\omega}{\kappa_{+}}\right)\left\{ f_{\omega lm}^{\mathrm{up}}\left(x\right),f_{\omega lm}^{\mathrm{up}*}\left(x'\right)\right\} +\left\{ f_{\omega lm}^{\mathrm{in}}\left(x\right),f_{\omega lm}^{\mathrm{in}*}\left(x'\right)\right\} \right],\label{G_U outside}
\end{equation}
where the subscript $U$ stands for ``Unruh state'', and the curly
brackets denote symmetrization with respect to the arguments $x$
and $x'$, i.e. 
\[
\left\{ A\left(x\right),B\left(x'\right)\right\} =A\left(x\right)B\left(x'\right)+A\left(x'\right)B\left(x\right).
\]
A similar procedure can be applied to the Hartle-Hawking Hadamard
function, yielding \cite{Candelas,Christensen =000026  Fulling}:
\begin{equation}
G_{H}^{\left(1\right)}\left(x,x'\right)=\intop_{0}^{\infty}d\omega\sum_{l,m}\coth\left(\frac{\pi\omega}{\kappa_{+}}\right)\left[\left\{ f_{\omega lm}^{\mathrm{up}}\left(x\right),f_{\omega lm}^{\mathrm{up}*}\left(x'\right)\right\} +\left\{ f_{\omega lm}^{\mathrm{in}}\left(x\right),f_{\omega lm}^{\mathrm{in}*}\left(x'\right)\right\} \right].\label{G_H outside}
\end{equation}
This expression may be obtained by replacing the above Unruh modes
with corresponding ``Hartle-Hawking modes'', which are Kruskal-based
at both $H_{\mathrm{past}}$ and PNI {[}more specifically, by changing
$e^{-i\omega v}\to e^{-i\omega V}$ at PNI in Eq. \eqref{g-in-up}{]}.

\section{The Unruh state Hadamard function inside the black hole \label{sec:The-Unruh-state}}

As discussed in Sec. \ref{Hadamard function outside}, outside the
BH the Unruh modes, and thereby the corresponding Hadamard function,
can be expressed in terms of the outer Eddington-Finkelstein modes.
These modes are naturally defined in the exterior region of the BH,
and they can be computed numerically by solving the ordinary differential
equation \eqref{radial-exterior} together with the boundary conditions
given by Eqs. \eqref{bc exterior in} and \eqref{bc exterior up}.
This way, using the relation between the former and the latter modes,
the Unruh state Hadamard function can be easily computed numerically. 

In the interior region, the Unruh state Hadamard function has the
same expression in terms of the Unruh modes as the one in the exterior,
since the Unruh modes are continuously defined throughout both regions
$I$ and $II$ of Fig. \ref{fig:Penrose diag RN} (as discussed in
Sec. \ref{Modes-and-quantum}). Using the same method as in the exterior
region, we can similarly obtain a relation between the Unruh modes
and the \emph{inner} Eddington-Finkelstein modes, which serve as the
internal analogues of the outer Eddington-Finkelstein modes. These
modes can be computed numerically by solving the ordinary differential
equation \eqref{radial-exterior} in the interior region together
with the initial condition given by Eq. \eqref{bdry cond}. Again,
this will facilitate the computation of the Unruh state Hadamard function
in the BH interior. 

We begin with the mode-sum expression for the Unruh state Hadamard
function in terms of the Unruh modes $g_{\omega lm}^{\mathrm{up}}$
and $g_{\omega lm}^{\mathrm{in}}$ : 

\begin{equation}
G_{U}^{\left(1\right)}\left(x,x'\right)=\left\langle \left\{ \hat{\Phi}\left(x\right),\hat{\Phi}\left(x'\right)\right\} \right\rangle _{U}=\intop_{0}^{\infty}d\omega\sum_{\Lambda,l,m}\left\{ g_{\omega lm}^{\Lambda}\left(x\right),g_{\omega lm}^{\mathrm{\Lambda}*}\left(x'\right)\right\} .\label{Unruh Hadamard fnc}
\end{equation}
Now, in order to express this function in the interior region in terms
of the inner Eddington-Finkelstein modes, we need to express the Unruh
modes using the latter. We first find the relation between these two
sets of modes on $H_{R}$ and $H_{L}$. For this purpose, we write
$\tilde{g}_{\omega l}^{\mathrm{past}}$ in terms of $\tilde{f}_{\omega l}^{\mathrm{up}}$
on $H_{\mathrm{past}}$, using Fourier transform, in the following
way:

\begin{equation}
\tilde{g}_{\omega l}^{\mathrm{past}}\left|_{H_{\mathrm{past}}}\right.\left(u_{\mathrm{ext}}\right)=e^{-i\omega U\left(u_{\mathrm{ext}}\right)}=\frac{1}{2\pi}\intop_{-\infty}^{\infty}\alpha_{\omega\tilde{\omega}}^{\mathrm{past}}e^{-i\tilde{\omega}u_{\mathrm{ext}}}d\tilde{\omega,}\label{g past}
\end{equation}
where $\alpha_{\omega\tilde{\omega}}^{\mathrm{past}}$ are the Fourier
coefficients given by the inverse Fourier transform
\[
\alpha_{\omega\tilde{\omega}}^{\mathrm{past}}=\intop_{-\infty}^{\infty}e^{-i\omega U\left(u_{\mathrm{ext}}\right)}e^{i\tilde{\omega}u_{\mathrm{ext}}}du_{\mathrm{ext}}.
\]
Substituting Eq. \eqref{Kruskal ext}, defining the variable $s=u_{\mathrm{ext}}-\frac{1}{\kappa_{+}}\log\left(\omega/\kappa_{+}\right)$
and using the identity \cite{Integral identity}
\[
\intop_{-\infty}^{\infty}e^{ie^{-\kappa_{+}s}}e^{i\tilde{\omega}s}ds=\frac{1}{\kappa_{+}}e^{\tilde{\omega}\pi/2\kappa_{+}}\varGamma\left(-i\frac{\tilde{\omega}}{\kappa_{+}}\right)\;,\;\mathrm{Im}\left(\tilde{\omega}\right)>0,
\]
we get (after assigning $\tilde{\omega}$ a small imaginary part and
taking it to vanish in the end) 

\begin{equation}
\begin{array}{c}
\alpha_{\omega\tilde{\omega}}^{\mathrm{past}}=\intop_{-\infty}^{\infty}e^{i\left(\omega/\kappa_{+}\right)e^{-\kappa_{+}u_{\mathrm{ext}}}}e^{i\tilde{\omega}u_{\mathrm{ext}}}du_{\mathrm{ext}}=\left(\frac{\omega}{\kappa_{+}}\right)^{i\tilde{\omega}/\kappa_{+}}\intop_{-\infty}^{\infty}e^{ie^{-\kappa_{+}s}}e^{i\tilde{\omega}s}ds\\
\\
=\frac{1}{\kappa_{+}}\left(\frac{\omega}{\kappa_{+}}\right)^{i\tilde{\omega}/\kappa_{+}}e^{\tilde{\omega}\pi/2\kappa_{+}}\varGamma\left(-i\frac{\tilde{\omega}}{\kappa_{+}}\right).
\end{array}\label{alpha past}
\end{equation}

Recalling that $\tilde{g}_{\omega l}^{\mathrm{past}}$ vanishes at
PNI, each mode $e^{-i\tilde{\omega}u_{\mathrm{ext}}}$ at $H_{\mathrm{past}}$
in Eq. \eqref{g past} evolves in time into a mode $\rho_{\tilde{\omega}l}^{\mathrm{up}}e^{-i\tilde{\omega}v}$
on $H_{R}$ (and an outgoing field $\tau_{\tilde{\omega}l}^{\mathrm{up}}e^{-i\tilde{\omega}u_{\mathrm{ext}}}$
at future null infinity, which will not concern us however). Thus
from linearity we get on $H_{R}$ 
\[
\left.\tilde{g}_{\omega l}^{\mathrm{past}}\right|_{H_{R}}=\frac{1}{2\pi}\intop_{-\infty}^{\infty}\alpha_{\omega\tilde{\omega}}^{\mathrm{past}}\rho_{\tilde{\omega}l}^{\mathrm{up}}e^{-i\tilde{\omega}v}d\tilde{\omega}=\frac{1}{2\pi}\intop_{-\infty}^{\infty}\alpha_{\omega\tilde{\omega}}^{\mathrm{past}}\rho_{\tilde{\omega}l}^{\mathrm{up}}\left.\tilde{f}_{\tilde{\omega}l}^{\mathrm{R}}\right|_{H_{R}}d\tilde{\omega}.
\]
Also, recalling that $\tilde{g}_{\omega l}^{\mathrm{past}}$ vanishes
on $H_{L}$ and so does $\tilde{f}_{\tilde{\omega}l}^{\mathrm{R}}$,
we get 
\[
\left.\tilde{g}_{\omega l}^{\mathrm{past}}\right|_{H_{L}}=0=\frac{1}{2\pi}\intop_{-\infty}^{\infty}\alpha_{\omega\tilde{\omega}}^{\mathrm{past}}\rho_{\tilde{\omega}l}^{\mathrm{up}}\left.\tilde{f}_{\tilde{\omega}l}^{\mathrm{R}}\right|_{H_{L}}d\tilde{\omega}.
\]
Since the wave equation \eqref{wave eq} is linear, these relations
between $\tilde{g}_{\omega l}^{\mathrm{past}}$ and $\tilde{f}_{\omega l}^{\mathrm{R}}$
will hold not only for points $x$ on $H_{L}$ and $H_{R}$, but also
for any point $x$ in the interior region of the BH. Therefore, throughout
region $II$ we can write 
\[
\tilde{g}_{\omega l}^{\mathrm{past}}\left(x\right)=\frac{1}{2\pi}\intop_{-\infty}^{\infty}\alpha_{\omega\tilde{\omega}}^{\mathrm{past}}\rho_{\tilde{\omega}l}^{\mathrm{up}}\tilde{f}_{\tilde{\omega}l}^{\mathrm{R}}\left(x\right)d\tilde{\omega}.
\]

In a similar fashion we now express $\tilde{g}_{\omega l}^{\mathrm{L}}$
in terms of $\tilde{f}_{\omega l}^{\mathrm{L}}$ on $H_{L}$ using
Fourier transform in the following way:

\begin{equation}
\tilde{g}_{\omega l}^{\mathrm{L}}\left|_{H_{L}}\right.\left(u_{\mathrm{int}}\right)=e^{-i\omega U\left(u_{\mathrm{int}}\right)}=\frac{1}{2\pi}\intop_{-\infty}^{\infty}\alpha_{\omega\tilde{\omega}}^{\mathrm{L}}e^{-i\tilde{\omega}u_{\mathrm{int}}}d\tilde{\omega}=\frac{1}{2\pi}\intop_{-\infty}^{\infty}\alpha_{\omega\tilde{\omega}}^{\mathrm{L}}\left.\tilde{f}_{\tilde{\omega}l}^{\mathrm{L}}\right|_{H_{L}}d\tilde{\omega},\label{g L}
\end{equation}
where again $\alpha_{\omega\tilde{\omega}}^{\mathrm{L}}$ are the
Fourier coefficients given by the inverse Fourier transform
\[
\alpha_{\omega\tilde{\omega}}^{\mathrm{L}}=\intop_{-\infty}^{\infty}e^{-i\omega U\left(u_{\mathrm{in}}\right)}e^{i\tilde{\omega}u_{\mathrm{in}}}du_{\mathrm{in}}.
\]
A similar computation to that shown in Eq. \eqref{alpha past} yields
\begin{equation}
\alpha_{\omega\tilde{\omega}}^{\mathrm{L}}=\frac{1}{\kappa_{+}}\left(\frac{\omega}{\kappa_{+}}\right)^{-i\tilde{\omega}/\kappa_{+}}e^{\tilde{\omega}\pi/2\kappa_{+}}\varGamma\left(i\frac{\tilde{\omega}}{\kappa_{+}}\right)=\alpha_{\omega\tilde{\omega}}^{\mathrm{past}*}.\label{alpha L}
\end{equation}
Using the same reasoning as above (recalling that $\tilde{g}_{\omega l}^{\mathrm{L}}$
and $\tilde{f}_{\omega l}^{\mathrm{L}}$ both vanish at $H_{R}$),
Eq. \eqref{g L} applies for a general point $x$ in the interior
region, therefore 
\[
\tilde{g}_{\omega l}^{\mathrm{L}}\left(x\right)=\frac{1}{2\pi}\intop_{-\infty}^{\infty}\alpha_{\omega\tilde{\omega}}^{\mathrm{L}}\tilde{f}_{\tilde{\omega}l}^{\mathrm{L}}\left(x\right)d\tilde{\omega}.
\]

We can now write the total ``up'' Unruh modes $\tilde{g}_{\omega l}^{\mathrm{up}}$
in the interior of the BH in terms of the inner Eddington-Finkelstein
modes as
\begin{equation}
\tilde{g}_{\omega l}^{\mathrm{up}}\left(x\right)=\tilde{g}_{\omega l}^{\mathrm{past}}\left(x\right)+\tilde{g}_{\omega l}^{\mathrm{L}}\left(x\right)=\frac{1}{2\pi}\intop_{-\infty}^{\infty}\left[\alpha_{\omega\tilde{\omega}}^{\mathrm{past}}\rho_{\tilde{\omega}l}^{\mathrm{up}}\tilde{f}_{\tilde{\omega}l}^{\mathrm{R}}\left(x\right)+\alpha_{\omega\tilde{\omega}}^{\mathrm{L}}\tilde{f}_{\tilde{\omega}l}^{\mathrm{L}}\left(x\right)\right]d\tilde{\omega}.\label{g tilda up}
\end{equation}
Using Eqs. \eqref{full Edding modes} and \eqref{full Unruh modes},
Eq. \eqref{g tilda up} can be readily written using the full modes
as
\begin{equation}
g_{\omega lm}^{\mathrm{up}}\left(x\right)=\frac{\left|\omega\right|^{-1/2}}{2\pi}\intop_{-\infty}^{\infty}\left[\alpha_{\omega\tilde{\omega}}^{\mathrm{past}}\rho_{\tilde{\omega}l}^{\mathrm{up}}f_{\tilde{\omega}lm}^{\mathrm{R}}\left(x\right)+\alpha_{\omega\tilde{\omega}}^{\mathrm{L}}f_{\tilde{\omega}lm}^{\mathrm{L}}\left(x\right)\right]\left|\tilde{\omega}\right|^{1/2}d\tilde{\omega}.\label{g up}
\end{equation}

We are now left with the ``in'' Unruh modes $\tilde{g}_{\omega l}^{\mathrm{in}}$.
Since the initial condition of these modes at PNI is given by $e^{-i\omega v}$,
the values of these modes on $H_{R}$ are $\tau_{\omega l}^{\mathrm{in}}e^{-i\omega v}$
(and they vanish on $H_{L}$). Recalling the initial conditions for
$\tilde{f}_{\omega l}^{\mathrm{R}}$, we find that for a general point
$x$ in the interior region
\begin{equation}
\tilde{g}_{\omega l}^{\mathrm{in}}\left(x\right)=\tau_{\omega l}^{\mathrm{in}}\tilde{f}_{\omega l}^{\mathrm{R}}\left(x\right).\label{g tilda in}
\end{equation}
Here as well, we can use Eqs. \eqref{full Edding modes} and \eqref{full Unruh modes}
and rewrite Eq. \eqref{g tilda in} as
\begin{equation}
g_{\omega lm}^{\mathrm{in}}\left(x\right)=\tau_{\omega l}^{\mathrm{in}}f_{\omega lm}^{\mathrm{R}}\left(x\right).\label{g in}
\end{equation}

We are now ready to compute Hadamard's function {[}Eq. \eqref{Hadamard}{]}
in the interior region of the BH in Unruh state, using Eqs. \eqref{Unruh Hadamard fnc},
\eqref{g up} and \eqref{g in}. It will be convenient to split the
Unruh state Hadamard function into a sum of two terms, one resulting
from the contribution of the ``up'' modes and the other from that
of the ``in'' modes: 
\[
G_{U}^{\left(1\right)}\left(x,x'\right)=G_{U}^{\left(1\right)\mathrm{up}}\left(x,x'\right)+G_{U}^{\left(1\right)\mathrm{in}}\left(x,x'\right).
\]
Let us first consider $G_{U}^{\left(1\right)\mathrm{up}}\left(x,x'\right)$.
Writing it as a mode sum, as in Eq. \eqref{Unruh Hadamard fnc}, we
have 
\[
G_{U}^{\left(1\right)\mathrm{up}}\left(x,x'\right)=\intop_{0}^{\infty}d\omega\sum_{l,m}\left\{ g_{\omega lm}^{\mathrm{up}}\left(x\right),g_{\omega lm}^{\mathrm{up}*}\left(x'\right)\right\} .
\]
Substituting Eq. \eqref{g up} in the above equation yields
\[
G_{U}^{\left(1\right)\mathrm{up}}\left(x,x'\right)=I_{RR}+I_{LL}+I_{RL}+I_{LR},
\]
where
\[
I_{RR}=\frac{1}{4\pi^{2}}\sum_{l,m}\intop_{-\infty}^{\infty}\left|\tilde{\omega}\right|^{1/2}d\tilde{\omega}\intop_{-\infty}^{\infty}\left|\tilde{\tilde{\omega}}\right|^{1/2}d\tilde{\tilde{\omega}}\rho_{\tilde{\omega}l}^{\mathrm{up}}\rho_{\tilde{\tilde{\omega}}l}^{\mathrm{up}*}\left\{ f_{\tilde{\omega}lm}^{\mathrm{R}}\left(x\right),f_{\tilde{\tilde{\omega}}lm}^{\mathrm{R}*}\left(x'\right)\right\} \intop_{0}^{\infty}\frac{d\omega}{\omega}\alpha_{\omega\tilde{\omega}}^{\mathrm{past}}\alpha_{\omega\tilde{\tilde{\omega}}}^{\mathrm{past}*},
\]
\[
I_{LL}=\frac{1}{4\pi^{2}}\sum_{l,m}\intop_{-\infty}^{\infty}\left|\tilde{\omega}\right|^{1/2}d\tilde{\omega}\intop_{-\infty}^{\infty}\left|\tilde{\tilde{\omega}}\right|^{1/2}d\tilde{\tilde{\omega}}\left\{ f_{\tilde{\omega}lm}^{\mathrm{L}}\left(x\right),f_{\tilde{\tilde{\omega}}lm}^{\mathrm{L}*}\left(x'\right)\right\} \intop_{0}^{\infty}\frac{d\omega}{\omega}\alpha_{\omega\tilde{\omega}}^{\mathrm{L}}\alpha_{\omega\tilde{\tilde{\omega}}}^{\mathrm{L}*},
\]

\[
I_{RL}=\frac{1}{4\pi^{2}}\sum_{l,m}\intop_{-\infty}^{\infty}\left|\tilde{\omega}\right|^{1/2}d\tilde{\omega}\intop_{-\infty}^{\infty}\left|\tilde{\tilde{\omega}}\right|^{1/2}d\tilde{\tilde{\omega}}\rho_{\tilde{\omega}l}^{\mathrm{up}}\left\{ f_{\tilde{\omega}lm}^{\mathrm{R}}\left(x\right),f_{\tilde{\tilde{\omega}}lm}^{\mathrm{L}*}\left(x'\right)\right\} \intop_{0}^{\infty}\frac{d\omega}{\omega}\alpha_{\omega\tilde{\omega}}^{\mathrm{past}}\alpha_{\omega\tilde{\tilde{\omega}}}^{\mathrm{L}*},
\]
\[
I_{LR}=\frac{1}{4\pi^{2}}\sum_{l,m}\intop_{-\infty}^{\infty}\left|\tilde{\omega}\right|^{1/2}d\tilde{\omega}\intop_{-\infty}^{\infty}\left|\tilde{\tilde{\omega}}\right|^{1/2}d\tilde{\tilde{\omega}}\rho_{\tilde{\tilde{\omega}}l}^{\mathrm{up}*}\left\{ f_{\tilde{\omega}lm}^{\mathrm{L}}\left(x\right),f_{\tilde{\tilde{\omega}}lm}^{\mathrm{R}*}\left(x'\right)\right\} \intop_{0}^{\infty}\frac{d\omega}{\omega}\alpha_{\omega\tilde{\omega}}^{\mathrm{L}}\alpha_{\omega\tilde{\tilde{\omega}}}^{\mathrm{past}*}=I_{RL}^{*}.
\]
Substituting Eqs. \eqref{alpha past} and \eqref{alpha L} in the
above expressions, and using the results
\[
\intop_{0}^{\infty}\frac{d\omega}{\omega}\alpha_{\omega\tilde{\omega}}^{\mathrm{past}}\alpha_{\omega\tilde{\tilde{\omega}}}^{\mathrm{past}*}=\intop_{0}^{\infty}\frac{d\omega}{\omega}\alpha_{\omega\tilde{\omega}}^{\mathrm{L}}\alpha_{\omega\tilde{\tilde{\omega}}}^{\mathrm{L}*}=\frac{4\pi^{2}}{\tilde{\omega}}\frac{1}{1-e^{-2\pi\tilde{\omega}/\kappa_{+}}}\delta\left(\tilde{\omega}-\tilde{\tilde{\omega}}\right)
\]
and
\[
\intop_{0}^{\infty}\frac{d\omega}{\omega}\alpha_{\omega\tilde{\omega}}^{\mathrm{past}}\alpha_{\omega\tilde{\tilde{\omega}}}^{\mathrm{L}*}=\intop_{0}^{\infty}\frac{d\omega}{\omega}\alpha_{\omega\tilde{\omega}}^{\mathrm{L}}\alpha_{\omega\tilde{\tilde{\omega}}}^{\mathrm{past}*}=\frac{2\pi^{2}}{\tilde{\omega}}\sinh^{-1}\left(\frac{\pi\omega}{\kappa_{+}}\right)\delta\left(\tilde{\omega}+\tilde{\tilde{\omega}}\right),
\]
we get (renaming the integration variable) 
\[
I_{RR}=\sum_{l,m}\intop_{-\infty}^{\infty}d\omega\,\mathrm{sgn}\left(\omega\right)\frac{1}{1-e^{-2\pi\omega/\kappa_{+}}}\left|\rho_{\omega l}^{\mathrm{up}}\right|^{2}\left\{ f_{\omega lm}^{\mathrm{R}}\left(x\right),f_{\omega lm}^{\mathrm{R}*}\left(x'\right)\right\} ,
\]
\[
I_{LL}=\sum_{l,m}\intop_{-\infty}^{\infty}d\omega\,\mathrm{sgn}\left(\omega\right)\frac{1}{1-e^{-2\pi\omega/\kappa_{+}}}\left\{ f_{\omega lm}^{\mathrm{L}}\left(x\right),f_{\omega lm}^{\mathrm{L}*}\left(x'\right)\right\} ,
\]
\[
I_{RL}=\frac{1}{2}\sum_{l,m}\intop_{-\infty}^{\infty}d\omega\,\mathrm{sgn}\left(\omega\right)\sinh^{-1}\left(\pi\frac{\omega}{\kappa_{+}}\right)\rho_{\omega l}^{\mathrm{up}}\left\{ f_{\omega lm}^{\mathrm{R}}\left(x\right),f_{\omega lm}^{\mathrm{L}*}\left(x'\right)\right\} ,
\]
\[
I_{LR}=I_{RL}^{*}.
\]
We now split each of these integrals into two integrals, one over
the positive values of $\omega$ and the other on the negative values.
We then use the identity
\[
C_{lm}\left(x\right)C_{lm}^{*}\left(x'\right)=C_{l\left(-m\right)}^{*}\left(x\right)C_{l\left(-m\right)}\left(x'\right),
\]
which further implies
\begin{equation}
\sum_{m=-l}^{l}C_{lm}\left(x\right)C_{lm}^{*}\left(x'\right)=\sum_{m=-l}^{l}C_{lm}^{*}\left(x\right)C_{lm}\left(x'\right),\label{identity sum cc}
\end{equation}
to obtain relations between the different modes, such as 
\[
\sum_{m=-l}^{l}f_{\left(-\omega\right)lm}^{R*}\left(x'\right)f_{\omega lm}^{L}\left(x\right)=\sum_{m=-l}^{l}f_{\left(-\omega\right)lm}^{L*}\left(x\right)f_{\omega lm}^{R}\left(x'\right).
\]
Using these relations and \eqref{identity sum cc} after summing the
four expressions for $I_{RR}$, $I_{LL}$, $I_{RL}$ and $I_{LR}$,
we finally get for the ``up'' part of Hadamard's function
\[
G_{U}^{\left(1\right)\mathrm{up}}\left(x,x'\right)=\intop_{0}^{\infty}d\omega\sum_{l,m}\left[\coth\left(\frac{\pi\omega}{\kappa_{+}}\right)\left(\left\{ f_{\omega lm}^{\mathrm{L}}\left(x\right),f_{\omega lm}^{\mathrm{L}*}\left(x'\right)\right\} +\left|\rho_{\omega l}^{\mathrm{up}}\right|^{2}\left\{ f_{\omega lm}^{\mathrm{R}}\left(x\right),f_{\omega lm}^{\mathrm{R}*}\left(x'\right)\right\} \right)\right.
\]
\begin{equation}
\left.+2\sinh^{-1}\left(\frac{\pi\omega}{\kappa_{+}}\right)\mathrm{Re}\left(\rho_{\omega l}^{\mathrm{up}}\left\{ f_{\omega lm}^{\mathrm{R}}\left(x\right),f_{\left(-\omega\right)lm}^{\mathrm{L}*}\left(x'\right)\right\} \right)\right].\label{Hadamard up}
\end{equation}

Next we consider the much simpler term $G_{U}^{\left(1\right)\mathrm{in}}\left(x,x'\right)$.
Writing it as a mode sum, we have
\[
G_{U}^{\left(1\right)\mathrm{in}}\left(x,x'\right)=\intop_{0}^{\infty}d\omega\sum_{l,m}\left\{ g_{\omega lm}^{\mathrm{in}}\left(x\right),g_{\omega lm}^{\mathrm{in}*}\left(x'\right)\right\} .
\]
Substituting Eq. \eqref{g in} in the above expression yields
\begin{equation}
G_{U}^{\left(1\right)\mathrm{in}}\left(x,x'\right)=\intop_{0}^{\infty}d\omega\sum_{l,m}\left|\tau_{\omega l}^{\mathrm{in}}\right|^{2}\left\{ f_{\omega lm}^{\mathrm{R}}\left(x\right),f_{\omega lm}^{\mathrm{R}*}\left(x'\right)\right\} .\label{Hadamard in}
\end{equation}
Summing Eqs. \eqref{Hadamard up} and \eqref{Hadamard in} gives the
final expression for Hadamard's function in Unruh state in terms of
the inner Eddington-Finkelstein modes in the interior region of the
BH. It is given by 
\[
G_{U}^{\left(1\right)}\left(x,x'\right)=G_{U}^{\left(1\right)\mathrm{up}}\left(x,x'\right)+G_{U}^{\left(1\right)\mathrm{in}}\left(x,x'\right)
\]
\[
=\intop_{0}^{\infty}d\omega\sum_{l,m}\left[\coth\left(\frac{\pi\omega}{\kappa_{+}}\right)\left\{ f_{\omega lm}^{\mathrm{L}}\left(x\right),f_{\omega lm}^{\mathrm{L}*}\left(x'\right)\right\} +\left(\coth\left(\frac{\pi\omega}{\kappa_{+}}\right)\left|\rho_{\omega l}^{\mathrm{up}}\right|^{2}+\left|\tau_{\omega l}^{\mathrm{up}}\right|^{2}\right)\left\{ f_{\omega lm}^{\mathrm{R}}\left(x\right),f_{\omega lm}^{\mathrm{R}*}\left(x'\right)\right\} \right.
\]
\begin{equation}
\left.+2\mathrm{csch}\left(\frac{\pi\omega}{\kappa_{+}}\right)\mathrm{Re}\left(\rho_{\omega l}^{\mathrm{up}}\left\{ f_{\omega lm}^{\mathrm{R}}\left(x\right),f_{\left(-\omega\right)lm}^{\mathrm{L}*}\left(x'\right)\right\} \right)\right],\label{Hadamard Unruh}
\end{equation}
where we used the relation
\begin{equation}
\left|\tau_{\omega l}^{\mathrm{up}}\right|=\left|\tau_{\omega l}^{\mathrm{in}}\right|.\label{tau in up}
\end{equation}

As discussed above in Sec. \ref{Modes-and-quantum}, the modes $f_{\omega lm}^{\mathrm{L}}$
and $f_{\omega lm}^{\mathrm{R}}$ can be obtained from Eqs. \eqref{full Edding modes}
and \eqref{decomp rad int} by numerically solving the radial equation
\eqref{radial-exterior} for $\psi_{\omega l}$, and can then be used
to construct $G_{U}^{\left(1\right)}$. 

\section{The Hartle-Hawking state Hadamard function inside the black hole
\label{sec:The-Hartle-Hawking-state}}

The Hartle-Hawking state, like the Unruh state, is regular across
the event horizon ($H_{R}$). In particular, quantities like the Hadamard
function and the renormalized stress-energy tensor should be analytic
\cite{Candelas_and_Jensen} across $r_{+}$. The expression for $G_{H}^{\left(1\right)}\left(x,x'\right)$
outside the BH is known, see Eq. \eqref{G_H outside}, and in principle
all we need is to analytically extend it from $r>r_{+}$ to $r<r_{+}$.
The main complication is that the functions $f_{\omega lm}^{\mathrm{up}}\left(x\right)$
are irregular at the event horizon (their asymptotic behavior is $\propto e^{-i\omega u_{\mathrm{ext}}}$,
which oscillates infinite times on approaching the even horizon),
making their analytic extension tricky.

To circumvent this difficulty we recall that since $G_{U}^{\left(1\right)}\left(x,x'\right)$
is regular at the event horizon too, the difference $G_{H}^{\left(1\right)}\left(x,x'\right)-G_{U}^{\left(1\right)}\left(x,x'\right)$
is also analytic at the event horizon. From Eqs. \eqref{G_U outside}
and \eqref{G_H outside} we obtain in the exterior region: 
\begin{equation}
G_{H}^{\left(1\right)}\left(x,x'\right)-G_{U}^{\left(1\right)}\left(x,x'\right)=\intop_{0}^{\infty}d\omega\sum_{l,m}\left[\coth\left(\frac{\pi\omega}{\kappa_{+}}\right)-1\right]\left\{ f_{\omega lm}^{\mathrm{in}}\left(x\right),f_{\omega lm}^{\mathrm{in}*}\left(x'\right)\right\} ,\qquad r>r_{+}.\label{eq:difference-Outside}
\end{equation}
It only involves the function $f_{\omega lm}^{\mathrm{in}}\left(x\right)$,
which is regular across the event horizon. This function is the solution
of the wave equation \eqref{KG} with boundary conditions $\propto e^{-i\omega v}$
at PNI and zero along $V=0$ (namely the union of $H_{\mathrm{past}}$
and $H_{L}$). This in itself guarantees the regularity of this function
at $r=r_{+}$, and also uniquely determines its extension to $r<r_{+}$.
It is convenient to describe this extension in terms of the associated
function $\tilde{f}_{\omega lm}^{\mathrm{in}}\left(x\right)$ (the
two functions are related by the trivial factor $\left|\omega\right|^{-1/2}C_{lm}$).
The asymptotic behavior of $\tilde{f}_{\omega lm}^{\mathrm{in}}$
is $e^{-i\omega v}$ at PNI and $\tau_{\omega l}^{\mathrm{in}}e^{-i\omega v}$
at $H_{R}$ (and zero at $V=0$). From Eq. \eqref{initial conditions L and R}
it immediately follows that its extension to $r<r_{+}$ is just $\tau_{\omega l}^{\mathrm{in}}\tilde{f}_{\omega lm}^{\mathrm{R}}$.
Likewise, the extension of $f_{\omega lm}^{\mathrm{in}}$ to $r<r_{+}$
is simply 
\[
f_{\omega lm}^{\mathrm{in}}\left(x\right)\to\tau_{\omega l}^{\mathrm{in}}f_{\omega lm}^{\mathrm{R}}\left(x\right)\,.
\]
Implementing this extension to Eq. \eqref{eq:difference-Outside}
yields the $H-U$ difference inside the BH: 
\begin{equation}
G_{H}^{\left(1\right)}\left(x,x'\right)-G_{U}^{\left(1\right)}\left(x,x'\right)=\intop_{0}^{\infty}d\omega\sum_{l,m}\left|\tau_{\omega l}^{\mathrm{up}}\right|^{2}\left[\coth\left(\frac{\pi\omega}{\kappa_{+}}\right)-1\right]\left\{ f_{\omega lm}^{\mathrm{R}}\left(x\right),f_{\omega lm}^{\mathrm{R}*}\left(x'\right)\right\} ,\qquad r<r_{+},\label{eq:difference-Inside}
\end{equation}
where again, we have used Eq. \eqref{tau in up}. Adding it to the
Unruh-state expression \eqref{Hadamard Unruh}, we finally obtain
the expression for the Hartle-Hawking state Hadamard function inside
the BH:

\[
G_{H}^{\left(1\right)}\left(x,x'\right)=\intop_{0}^{\infty}d\omega\sum_{l,m}\left[\coth\left(\frac{\pi\omega}{\kappa_{+}}\right)\left(\left\{ f_{\omega lm}^{\mathrm{L}}\left(x\right),f_{\omega lm}^{\mathrm{L}*}\left(x'\right)\right\} +\left\{ f_{\omega lm}^{\mathrm{R}}\left(x\right),f_{\omega lm}^{\mathrm{R}*}\left(x'\right)\right\} \right)\right.
\]
\begin{equation}
\left.+2\mathrm{csch}\left(\frac{\pi\omega}{\kappa_{+}}\right)\mathrm{Re}\left(\rho_{\omega l}^{\mathrm{up}}\left\{ f_{\omega lm}^{\mathrm{R}}\left(x\right),f_{\left(-\omega\right)lm}^{\mathrm{L}*}\left(x'\right)\right\} \right)\right],\label{Hadamard H-H}
\end{equation}
where we have used the relation
\[
\left|\rho_{\omega l}^{\mathrm{up}}\right|^{2}+\left|\tau_{\omega l}^{\mathrm{up}}\right|^{2}=1.
\]

Again, as discussed in the previous section for the Unruh state, $G_{H}^{\left(1\right)}$
can be expressed in terms of the radial function $\psi_{\omega l}$,
which can be computed numerically from the radial equation \eqref{radial-exterior}
together with the boundary condition \eqref{bdry cond}.

\section{Trace of the RSET for a minimally-coupled scalar field \label{sec:trace-anomaly-for}}

In this section we derive a simple expression for the trace $<T_{\alpha}^{\alpha}>_{ren}$
of a minimally-coupled massless scalar field. Such an explicit expression
turns out to be useful for RSET analysis in curved spacetime (particularly
inside BHs). 

For a conformally-coupled massless scalar field the trace $T_{\alpha}^{\alpha}$
of the classical energy-momentum tensor strictly vanishes. As a consequence,
at the quantum level the renormalized expectation value $<T_{\alpha}^{\alpha}>{}_{ren}$
becomes a purely local quantity (independent of the quantum state).
This is the well-known trace anomaly \cite{Trace anomaly 1}: 
\begin{equation}
\left\langle T_{\alpha}^{\alpha}\right\rangle {}_{ren}=T_{\mathrm{anomaly}}\qquad\qquad\qquad\mbox{(conformal field)}\label{eq:trace_conformal}
\end{equation}
where 
\[
T_{\mathrm{anomaly}}\equiv\frac{1}{2880\pi^{2}}\left(R_{\alpha\beta\gamma\delta}R^{\alpha\beta\gamma\delta}-R_{\alpha\beta}R^{\alpha\beta}+\frac{5}{2}R^{2}+6\boxempty R\right).
\]

For a minimally coupled scalar field $\phi$ (which is not conformally
coupled in 4D), this expression no longer holds, because the classical
trace $T_{\mu}^{\mu}$ does not vanish. Nevertheless, a simple generalization
of Eq. \eqref{eq:trace_conformal} still holds as we now show. 

In the minimally-coupled massless case the classical stress-energy
tensor is 
\begin{equation}
T_{\mu\nu}=\mbox{\ensuremath{\phi}}_{;\mu}\mbox{\ensuremath{\phi}}_{;\nu}-\frac{1}{2}g_{\mu\nu}\phi^{;\alpha}\mbox{\ensuremath{\phi}}_{;\alpha}\,,\label{eq:classical_stress}
\end{equation}
hence
\begin{equation}
T_{\alpha}^{\alpha}=-\phi^{;\alpha}\mbox{\ensuremath{\phi}}_{;\alpha}\,\label{eq:classical_trace}
\end{equation}
which is non-vanishing. To this expression we now add the quantity
$\boxempty\left(\phi^{2}\right)\cdot\mathrm{const}$. The field equation
$\boxempty\phi=0$ implies 

\[
\boxempty\left(\phi^{2}\right)=2\phi^{;\alpha}\mbox{\ensuremath{\phi}}_{;\alpha}\,.
\]
Correspondingly, we set $\mathrm{const}=1/2$ and obtain the classical
relation 
\[
T\equiv T_{\alpha}^{\alpha}+\frac{1}{2}\boxempty\left(\phi^{2}\right)=0\,.
\]
It then follows that in quantum field theory $\left\langle T\right\rangle {}_{ren}$
must be purely local. Namely,
\begin{equation}
\left\langle T_{\alpha}^{\alpha}\right\rangle {}_{ren}+\frac{1}{2}\boxempty\left\langle \hat{\Phi}^{2}\right\rangle _{ren}=T_{\mathrm{loc}}\,,\label{eq:intermed}
\end{equation}
where $T_{\mathrm{loc}}$ is some local geometric quantity. 

The explicit form of $T_{\mathrm{loc}}$ may be obtained from the
local (Hadamard-based) short-distance asymptotic behavior of $T_{\alpha}^{\alpha}$.
Such a local analysis was carried out by Brown \& Ottewill \cite{Trace anomaly 2},
see in particular Eq. (2.24) therein \footnote{In this equation, $w_{A}/(8\pi^{2})$ should coincide with $<\phi^{2}>_{ren}$.
We are grateful to A. Ottewill for pointing this out to us.} (setting $m=\xi=0$). One readily finds that $T_{\mathrm{loc}}$
is just the usual trace-anomaly term. Our final result is thus 
\begin{equation}
\left\langle T_{\alpha}^{\alpha}\right\rangle {}_{ren}=T_{\mathrm{anomaly}}-\frac{1}{2}\boxempty\left\langle \hat{\Phi}^{2}\right\rangle _{ren}\qquad\mbox{(minimally coupled massless field)}\,.\label{eq:generalized_anomaly}
\end{equation}

The procedure of adding a surface term to the stress-energy tensor
(to form a new tensor with a vanishing trace) is basically well known.
\footnote{An example of such operations with surface terms is given in \cite{ref 3}}
Here we used a similar idea to obtain the simple explicit relation
\eqref{eq:generalized_anomaly} between the two quantities $<T_{\alpha}^{\alpha}>_{ren}$
and $<\hat{\Phi}^{2}>_{ren}$ that are routinely computed in semiclassical
calculations. 

\section{Discussion\label{sec:Discussion}}

In this paper we developed explicit expressions for the Hadamard function
on the interior part of a RN black hole, using the internal Eddington-Finkelstein
modes, in both the Unruh and Hartle-Hawking states. Although we do
not present numerical results here, this scheme was recently used
\cite{Assaf-Adam-Amos-Preparation} to reproduce the results by Candelas
and Jensen \cite{Candelas_and_Jensen} for $<\hat{\Phi}^{2}>_{ren}$
in the interior of a Schwarzschild BH in the Hartle-Hawking state. 

This infrastructure is part of an ongoing effort to study the RSET
inside a RN black-hole, both analytically and numerically, with special
emphasis on the asymptotic behavior on approaching the Cauchy horizon. 

As a by-product of the above research we encountered the relation
\eqref{eq:generalized_anomaly} between the trace of the RSET and
the d'Alembertian of $<\hat{\Phi}^{2}>_{ren}$ for a minimally-coupled
massless scalar field. The identity reveals another piece of the puzzle
and also allows to check the results obtained for $<\hat{\Phi}^{2}>_{ren}$
against those obtained for the RSET. 

The foundations laid here, together with the PMR method, allow for
studying the quantum effects in the interior of charged BHs. We view
this as a first step towards the more important and more ambitious
goal of studying quantum effects inside rotating BHs, which are of
course much more realistic. 
\begin{acknowledgments}
We are grateful to Robert Wald, Paul Anderson, and Adrian Ottewill
for helpful discussions. This research was supported by the Asher
Fund for Space Research at the Technion.
\end{acknowledgments}

\end{document}